\documentstyle[mprocl]{article}

\def\be{\begin{equation}}
\def\ee{\end{equation}}
\def\bea{\begin{eqnarray}}
\def\eea{\end{eqnarray}}

\begin{document}

\title{The Mixmaster Universe is Unambiguously Chaotic}

\author{ N.J. CORNISH }

\address{DAMTP, Cambridge University, Silver Street, Cambridge, CB3
9EW, UK}

\author{ J.J. LEVIN}

\address{CfPA, University of California at Berkeley, 301 Le Conte
Hall, Berkeley, CA 94720-7304, USA}


\maketitle\abstracts{The Mixmaster or Bianchi IX cosmological model
has become one of the archetypal settings for studying gravitational
dynamics. The past decade has seen a vigourous debate about
whether or not the Mixmaster's dynamics is chaotic. In this talk we
review our recent work in uncovering a chaotic invariant set of orbits
in the Mixmaster phase space, and how we used this discovery to prove
the dynamics is chaotic.
}
  
\section{The mixmaster universe}

The mixmaster model\cite{mis,bkl} describes a homogeneous but
anisotropic closed
cosmology of Bianchi type IX. The spatial sections have the geometry
of a deformed three-sphere, evolving with three independent scale factors,
$a(t), b(t), c(t)$. The vacuum field equations read:
\be
(\ln a^2)''=(b^2-c^2)^2-a^4 \quad {\it et\; cyc.\; } (a,b,c) \, .
\label{fe}
\ee
Here a prime denotes $d/d\tau$ and $dt=abc\; d\tau$.
When the right hand sides of the three equations (\ref{fe})
are all $\approx 0$, {\it ie.} when the minisuperspace potential
\be
V= a^4+b^4+c^4-2(a^2b^2+a^2c^2+b^2c^2)\approx 0 \, , 
\ee
the universe evolves in an approximate Kasner phase with
$a\sim t^{p_{i}}, \; b \sim t^{p_{j}}, \; c \sim t^{p_{k}}$.
The Kasner exponents satisfy $\sum_{i=1}^{3} p_{i} =\sum_{i=1}^{3}
p_{i}^2=1$.

The dynamics of the model can be separated into the evolution of an
overall scale factor $\Omega$, and two variables that measure the
departure from isotropy\cite{mis}:
\be
\Omega=\frac{1}{3}\ln (abc),\quad \beta_+=\frac{1}{3}\ln(bc/a^2),
\quad \beta_-=\frac{1}{\sqrt{3}}\ln(b/c) \, .
\ee
Since it can be shown\cite{wald} that the evolution of $\Omega$
describes a smooth overall expansion and contraction of the universe,
any chaotic behaviour must be associated with motion in the anisotropy
plane $(\beta_+,\beta_-)$.

Early work on the mixmaster dynamics focused on discrete maps\cite{bkl} that
were used to approximate the continuum dynamics. The maps were shown
to be chaotic\cite{bkl,barrow}, and from this it was inferred that the
full dynamics was also chaotic. The picture became clouded in the late
eighties when numerical studies of the dynamics found that
the principle Lyapunov exponent vanished\cite{fm}, indicating that the
system was not chaotic. Subsequently, two possible causes for the
discrepancy were isolated\cite{hob}: (1) the full dynamics might be
integrable, but the approximate maps fail to preserve all the constants
of motion; (2) the full dynamics is chaotic but the Lyapunov exponents vanish
for certain choices of time coordinate. The second possibility raised
the important issue of how to describe chaos in a coordinate invariant
way. We address this question in our other contribution to this
volume. In the remainder of this talk we describe how coordinate
independent methods can be used to settle the mixmaster debate in
favour of option (2). That is, the mixmaster universe is chaotic and
the approximate maps do capture the essential features of the dynamics.

\section{Uncovering the chaotic invariant set}

The key to understanding the mixmaster dynamics can be found in the
very first papers on the subject\cite{mis,bkl}. These papers
identify three ``attractors'' for the dynamics corresponding to the
three states of highly anisotropic expansion $a\sim b\sim {\rm
const.}, \; c\sim t$, $c\sim a\sim {\rm
const.}, \; b\sim t$ and $b\sim c\sim {\rm const.}, \; a\sim t$. The
$SO(3)$ symmetry of the spatial sections ensures that all three
possibilities occur with equal probability. If these attractors
corresponded to final outcomes, 
it would be a simple matter to decide if the dynamics was chaotic. All
one would have to do is assign a colour to each outcome, then
colour-code the space of initial conditions according to the outcomes. If the
boundaries between the three outcomes are fractal, then the dynamics
is chaotic.

\begin{figure}[h]
\vspace*{2.3in}
\includegraphics{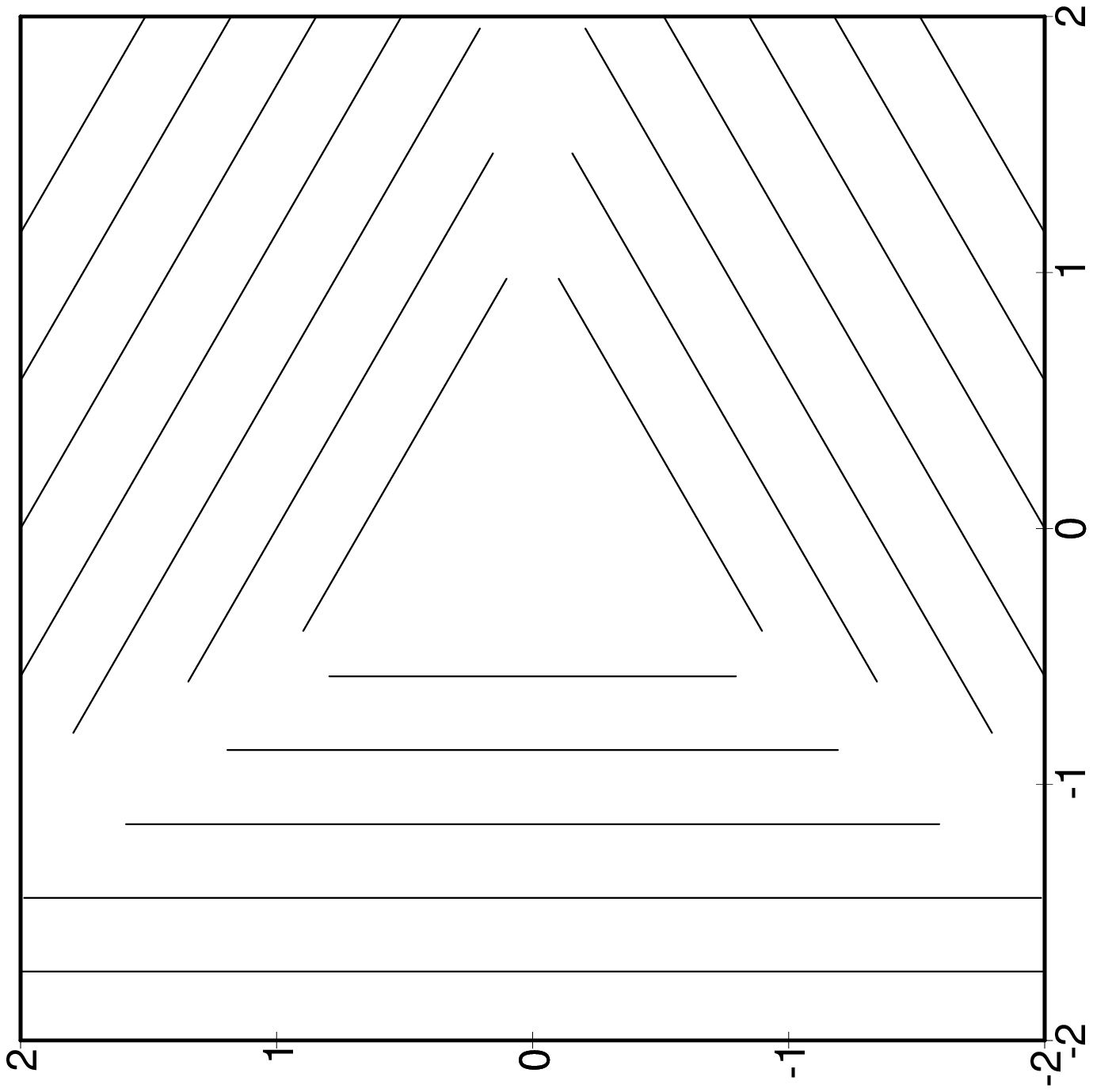}
\includegraphics{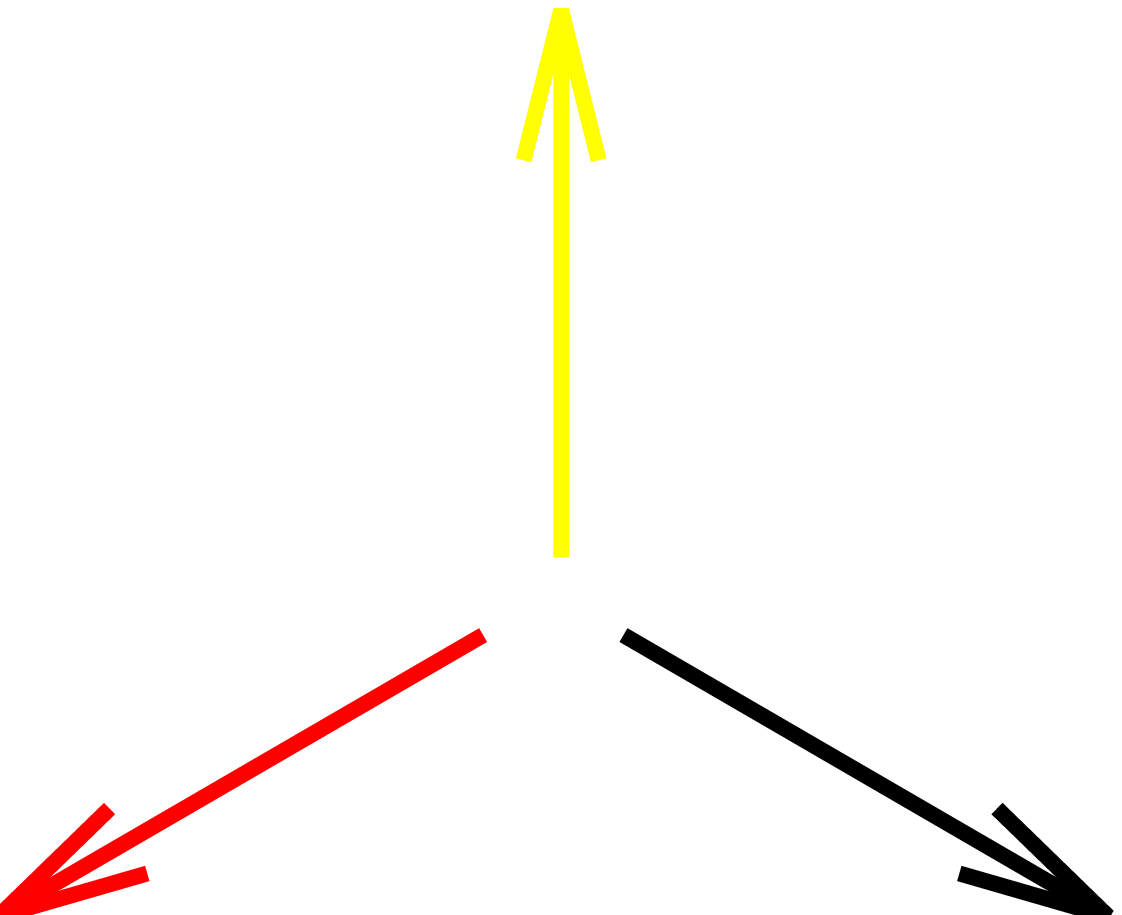}
\includegraphics{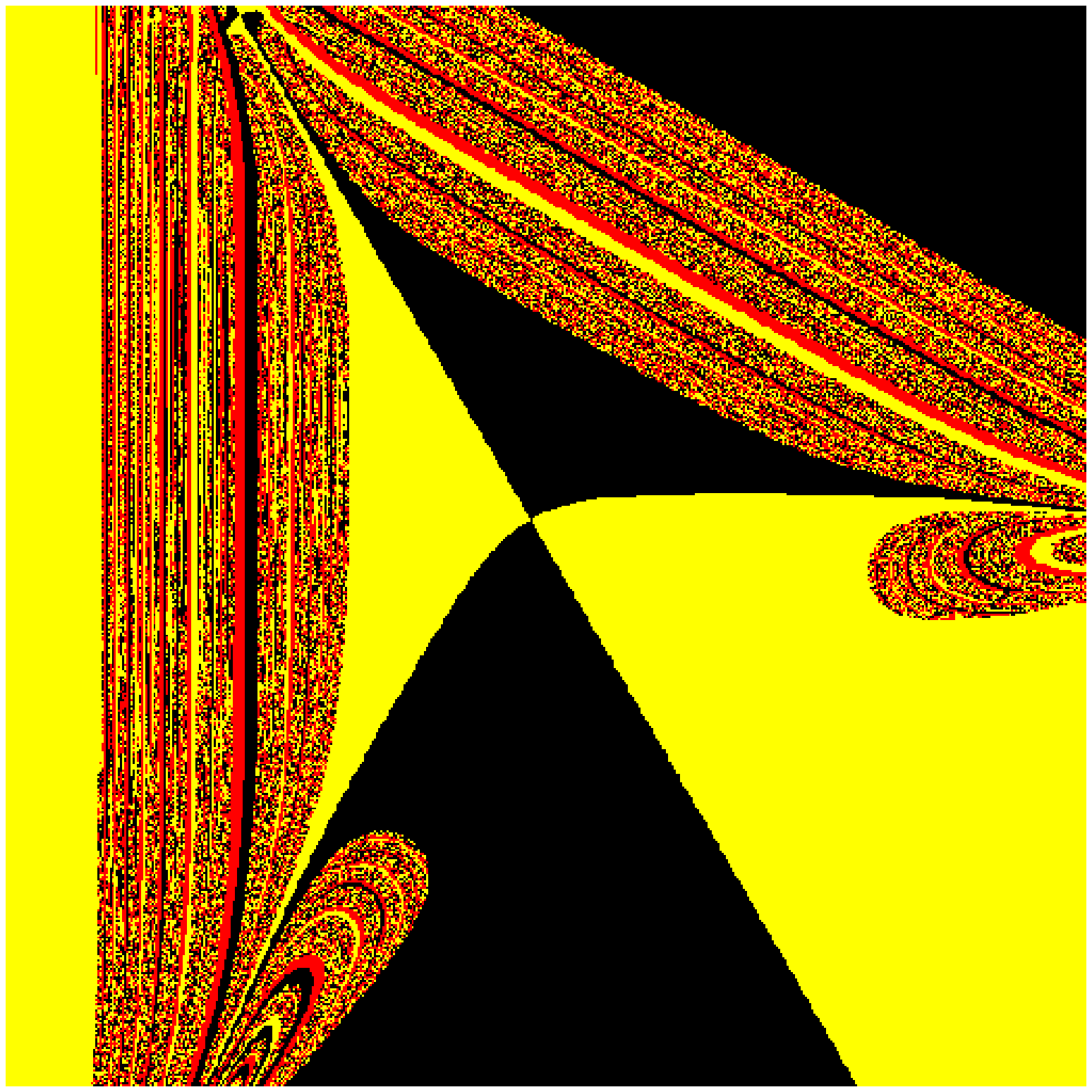}
\caption{(i) The minisuperspace potential with exits cut in the
corners and outcomes marked by arrows.
(ii) Outcome basins and their fractal boundaries in
the anisotropy plane.}
\end{figure}
\vspace*{-0.2in}
\noindent\begin{picture}(0,0)
\put(2,127){${ \beta_{-}}$}
\put(94,39){${\beta_{+}}$}
\put(187,127){${ \beta_{-}}$}
\put(275,40){${\beta_{+} }$}
\put(5,195){(i)}
\put(185,195){(ii)}
\put(47,125){{\Large {\bf y}}}
\put(115,82){{\Large {\bf z}}}
\put(115,167){{\Large {\bf x}}}
\end{picture}

Unfortunately, the procedure is not so simple as the attractors do not
correspond to final outcomes. When viewed in the anisotropy plane, the
attractors correspond to trajectories that go deep into one of the
three corners of the minisuperspace potential (see
Fig.(1.i)). After a long time, the trajectory returns to the scattering
region close to the centre of the triangular potential before again
scattering toward one of the three attractors.

We reasoned that if the mixmaster was chaotic, it must correspond to a
non-compact billiard\cite{cl}. Moreover, there should be a chaotic invariant
set of unstable periodic orbits in the centre of the scattering
region. In order to uncover this set we set a threshold based on the
relative rate of expansion of the three scale factors. When one axis
was collapsing very much faster than the other two we stopped the
evolution and assigned an outcome. This is equivalent to cutting
pockets in the corners of the minisuperspace potential, as shown in
Fig.(1.i). In effect we simplified the analysis by converting the mixmaster
into a Hamiltonian exit system\cite{blh}. We can do this with
confidence since a system with exits is always {\em less} chaotic than
the corresponding system without exits. 

With exits in place it is a simple matter to numerically plot the
attractor basin boundaries. A representative slice through phase space
is shown in Fig.(1.ii). The basin boundaries are clearly fractal, and
were found to have an information dimension\cite{cl} of $D_1=1.86\pm
0.02$. From this we can conclude in a coordinate independent way that
the mixmaster universe is chaotic. Moreover, exactly the same
procedure can be used to show that the Bianchi VIII cosmological
model is chaotic.

Trajectories belonging to the boundaries of the attractor basins are
future asymptotic to the unstable periodic orbits that comprise the
mixmaster's chaotic invariant set. (There are actually no truly periodic
orbits in the $(\beta_+,\beta_-)$ plane since the overall anisotropy
increases as a cosmological singularity is approached. The orbits are
self-similar rather than periodic. However, rescaled
coordinates can be found where the orbits are periodic). The
complexity of the mixmaster system can be measured by introducing a
symbolic coding for the periodic orbits. (A similar coding was
introduced by Rugh\cite{hob} for the aperiodic orbits).
The most efficient way of
coding an orbit is shown in Fig.(1.i). The symbol {\bf x}, {\bf y}, or
{\bf z} is recorded each time the trajectory passes through one
of the three triangular regions bounded by the corners of the
potential. The simplest periodic orbit, first found by
Misner\cite{mis}, corresponds to the repeated sequence ${\bf xyz}$.
All the periodic orbits can be similarly described,
and it is easy to show that the number of orbits made from
words of length $k$ grows as $N(k)\sim 2^k$. 
Since the number of orbits grows exponentially with length, the
collection of periodic orbits has a non-zero topological entropy, in
this case, $H_T=\ln 2$, which implies the dynamics is chaotic.

\section*{Acknowledgments}
We are grateful to Norm Frankel for his encouragement and enthusiasm.

\end{document}